# Experimental study of radiative shocks at PALS facility.

Chantal Stehlé (1), Matthias González (2,3),

Michaela Kozlova (4), Bedrich Rus (4), Tomas Mocek (4),

Ouali Acef (5), Jean Philippe Colombier (1, 6), Thierry Lanz (1, 7),

Norbert Champion (1), Krzysztof Jakubczak (4), Jiri Polan (4),

Patrice Barroso (8), Daniel Bauduin (8), Edouard Audit (3), Jan Dostal (4), Michal Stupka (4)

[1] LERMA, UMR 8112, Observatoire de Paris, CNRS et UPMC, 5 place J. Janssen, 92195 Meudon, France

[2] Instituto de Fusión Nuclear, Universidad Politécnica de Madrid, Madrid, Spain

[2] Laboratoire AIM, CEA/DSM/IRFU-CNRS-Université Paris Diderot, F-91191 Gif-sur-Yvette Cedex, France

[4] Department of X-Ray Lasers, Institute of Physics / PALS Center, Prague 8, Czech Republic

[5] SYRTE, UMR 8630, Observatoire de Paris, CNRS et Université Pierre et Marie Curie, 77 Avenue Denfert Rochereau, 75014 Paris, France

[6] Laboratoire Hubert Curien, UMR 5516, CNRS, et Université de Saint Etienne ,18 Rue du Professeur Benoît Lauras, 42000 Saint-Etienne, France

[7] Department of Astronomy, University of Maryland, College Park, MD 20742, USA

[8] GEPI, UMR 8111, Observatoire de Paris, CNRS et Université Diderot, 77 Avenue Denfert Rochereau, 75014 Paris, France

**Corresponding author** : Chantal Stehlé, Observatoire de Paris, LERMA, 5 Place Jules Janssen, 92195 Meudon France +33 1 45 07 74 16  mail : chantal.stehle@obspm.fr

**Short title** : Experimental Radiative Shocks

**Number of manuscript pages, including figures : 21**

**Number of tables  :  1**

**Number of figures :  7**

# Experimental study of radiative shocks at PALS facility.

## Abstract

We report on the investigation of strong radiative shocks generated with the high energy, sub-nanosecond iodine laser at PALS. These shock waves are characterized by a developed radiative precursor and their dynamics is analyzed over long time scales (~50 ns), approaching a quasi-stationary limit. We present the first preliminary results on the rear side XUV spectroscopy. These studies are relevant to the understanding of the spectroscopic signatures of accretion shocks in Classical T Tauri Stars.

**Keywords :** laboratory astrophysics - hydrodynamics - laser plasmas - shocks - stellar formation

# 1 – INTRODUCTION

Classical T Tauri stars are young optically visible low-mass stars surrounded by an accretion disk, undergoing accretion and outflow processes (Bertout 1989). Gas falls from the circumstellar envelope onto the stellar photosphere at a velocity of several hundreds of km/s, funnelled in accretion columns by the magnetic field (Bouvier et al. 2007). This accretion process engenders strong shocks and results in an excess of flux relative to the stellar photospheric flux in the optical and the ultraviolet domains. Accretion is also connected to the indirectly observed hot spots at the stellar surface (Donati et al. 2008). The dynamical and radiative properties of these accretion shocks remains an outstanding issue. Taking into account the temperature and density of the stellar photosphere together with the free fall velocity, the generated shocks may be in the regime of supercritical shocks with a developed radiative precursor. The physical structure of such radiative shocks is complex, as a consequence of the interplay between radiation and hydrodynamics, and needs to be characterized. This task can now be undertaken using high-energy laser installations, like the Prague PALS laser facility, which are able to generate radiative shocks in gases at velocities of about 60 km/s in tubes of millimetric scale (Gonzalez et al. 2006, Busquet et al. 2007).

Radiative shocks are characterized by an ionization front induced by the shock wave (Zeldovich & Raiser 1966, Kiselev et al. 1991). For a given shock velocity, which is set by the available laser energy, and a given initial gas pressure, these radiative effects are more important for species with high atomic number like Xenon (Bozier et al. 1986, Fleury et al. 2002, Bouquet et al. 2004, Reighard et al. 2005, Reighard et al. 2007, Rodriguez et al. 2008). These experimental studies have shown the importance of radiative losses at the tube walls, decreasing the velocity of the ionisation front. We have also shown theoretically that these lateral radiation losses result in a faster convergence of the shock structure toward the stationary limit (Gonzalez et al. 2009).

This effect may be similarly important in the accretion funnels of young forming stars, because of their limited lateral extension. Studying the physical and spectroscopic properties of radiative shocks in such a geometry in an experimental setting with controlled initial conditions may therefore be an ideal way for understanding the physics of accretion funnels. In this paper, we present experimental results of such shock waves over long times (~ 50 ns), aiming to reach the stationary limit discussed by Gonzalez et al. (2009) and to investigate the feasibility of back side time integrated XUV spectroscopy.

## 2 – EXPERIMENTAL SETUP

The experiment was carried out during the Fall of 2007 at the Prague Asterix Laser System (PALS) (Jungwirth et al. 2001, Batani et al. 2007; Kasperczuk et al. 2009) with an iodine laser ($\lambda_1$= 1.315 μm). The experimental design (Figure 1) includes two main diagnostics: time-resolved interferometry using a green laser (527 nm) to probe the electron density in the radiative precursor, and time-integrated back side XUV spectroscopy. The main blue laser pulse of the PALS installation, working at $\lambda_3$= 438 nm, is focussed on the target which consists of a miniaturized shock tube located in the centre of the vacuum chamber. The nominal pulse condition of the main PALS laser was an energy of 150 J per pulse lasting 0.35 ns.

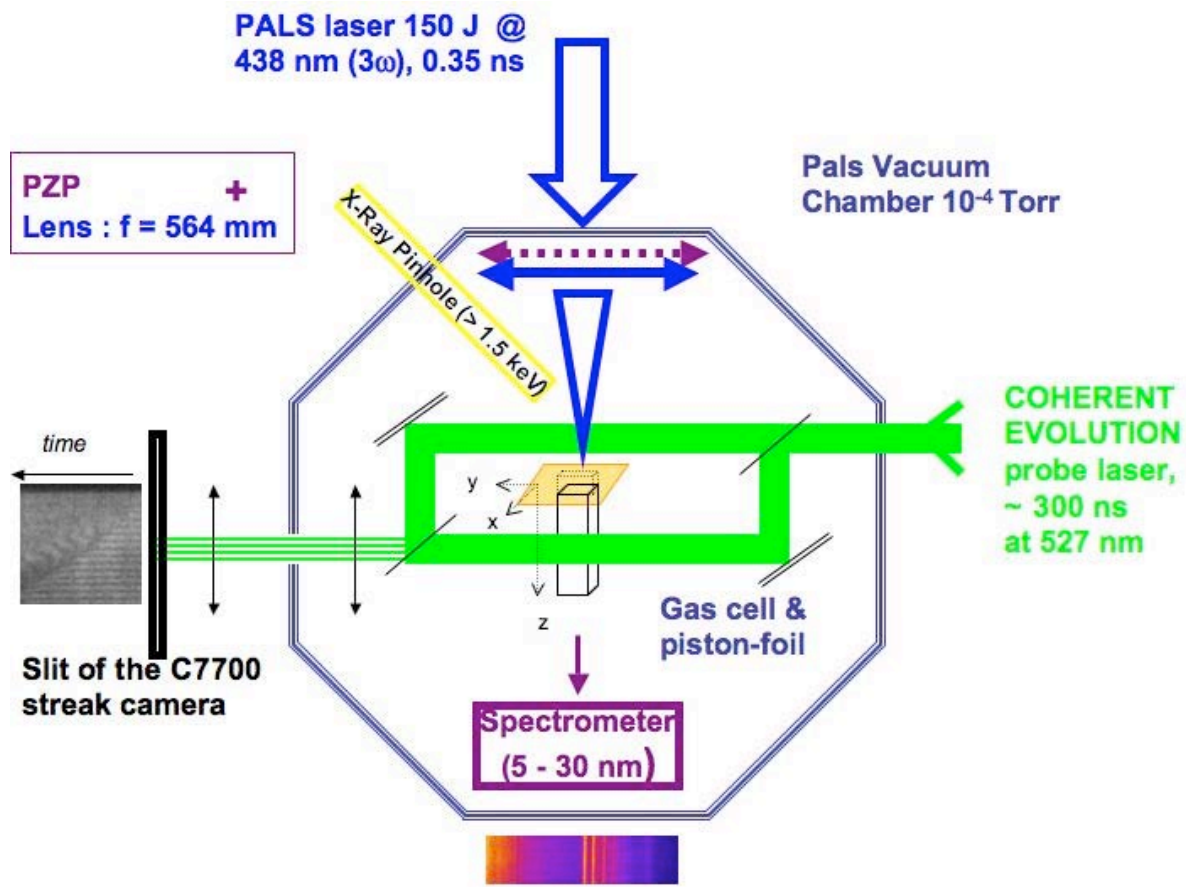

Figure 1: Experimental setup (see Paragraph 2 for details).

The PALS beam with a diameter of 290 mm is smoothed by a ColsiCoat Phase Zone Plate (PZP), located behind the main PALS aspherical lens (f = 564 mm at $\lambda_3$=3/$\lambda_1$, diameter d

= 290 mm). The PZP is designed to obtain a theoretical focal spot of diameter ~ 0.55 mm on the target, matching the inner section of the smallest shock tubes, where intensities of 1 - 2 x $10^{14}$ W/cm$^2$ are obtained for an energy pulse of ~ 150 J. The beam homogeneity on the target is controlled by a time-integrated X-ray pinhole camera imaging system, with a magnification of unity, through a 50 μm aluminium filter.

For the time-resolved interferometry diagnostic, we use an auxiliary (Coherent Evolution 15) laser working at $\lambda_P$=527 nm, with a pulse duration of ~ 300 ns. The target is located in one arm of a Mach–Zehnder interferometer. After recombination, the two laser beams of the interferometer are imaged on the slit of a C7700 HAMAMATSU streak camera. An interferometric filter at $\lambda_3$ and coloured glasses, located in front of the slit, remove spurious contributions from the plasma self-emission or from the scattered light of the PALS laser. Two configurations are used: *(i)* in the longitudinal interferometry, we image the symmetry axis (z) of the shock tube along the slit of the camera in order to monitor the shock propagation in the tube; *(ii)* in the transverse interferometry, we image a section of the tube, perpendicular (x) to the canal centre, to monitor the time variations of the transverse shape of the shock wave (see Figure 1 for axis orientation), at some fixed distance $z_0$ from the piston initial position.

The XUV time-integrated spectroscopic diagnostic consists of a grating spectrometer working between 5 and 30 nm, using a cylindrical gold coated grating (curvature radius of 5649 mm, 1200 groves per mm, efficient grating area of 45 x 27 mm, blaze angle 3.7°) placed at about 250 mm from the back side of the target. The spectral resolution is about 0.25 nm and the collection angle ~ 1.4 $10^{-4}$ rad. The signal is filtered with a 1.6 μm aluminium foil and recorded on a cooled ANDOR DX 440 CCD, located at a distance of 237 mm from the grating. The wavelength calibration has been performed with carbon lines from polypropylene targets and the aluminium L edge (at $\lambda$ = 17.1 nm) of aluminium foils.

The targets were manufactured at the Pôle Instrumental of Observatoire de Paris. We use miniaturized shock tubes filled with Xenon at low pressure with two generic geometries: *(a)* Aluminium parallelepiped tubes with a square internal section, two metallic sides and two lateral glass windows (Figure 2a) allowing for interferometric studies and *(b)* blind cylindrical metallic tubes without any lateral window (Figure 2b) for backside XUV spectroscopy in the direction of shock propagation.

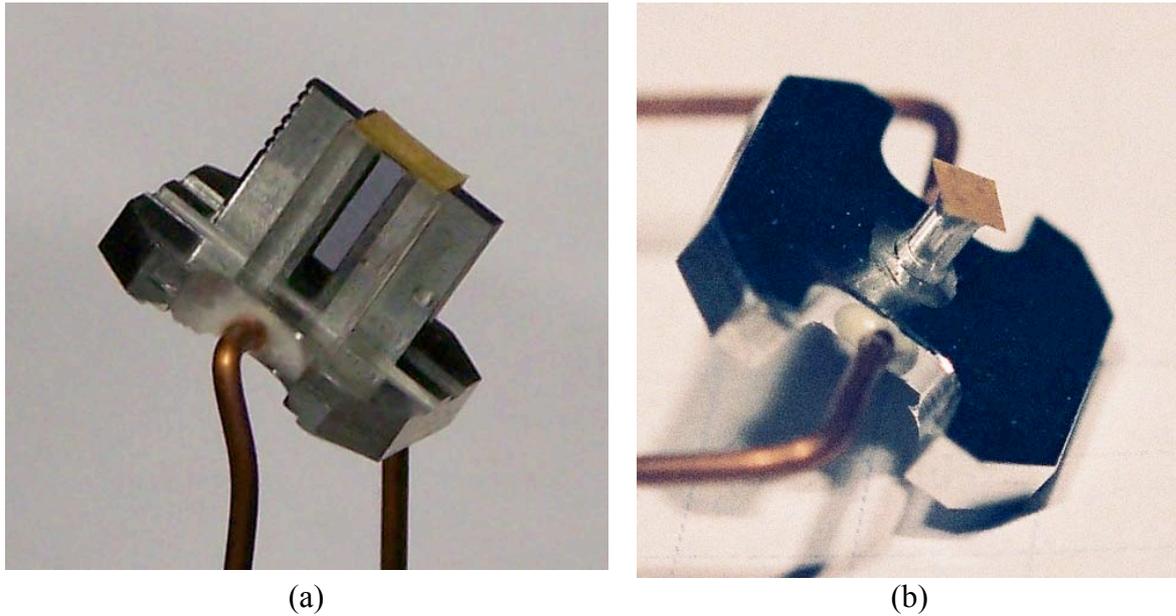

(a)                      (b)

Figure 2: Parallelepiped target with lateral glass windows *(a)* and blind cylindrical target *(b)* *(courtesy of Observatoire de Paris)*.

For the interferometric diagnostic, we use two different targets. The first target, F7, consists of a gilt aluminium, 6 mm long and U-shaped tube closed laterally by two BK7 glass windows with internal gold coating (15 nm thick). These windows allow for interferometric diagnostics with the auxiliary green laser. The internal section of this square tube is 0.7 x 0.7 mm$^2$. The drilled tube socket (2 mm thick), also in gilt aluminium, is used for gas filling through capillaries and is closed at the back by an "XUV window" for backside XUV spectroscopic study. The second square target, F61, is similar to F7, but without the gold coating. The internal section of the tube is larger (1.2 x 1.2 mm$^2$), and the internal faces of the BK7 windows have an aluminium coating (5 nm thick). The shock thus propagates in an aluminium channel for the F61 target, whereas it is a gilt channel in the first case (F7). The nature of the channel coating affects the albedo of the tube walls, thus the radiative properties of the shock, and must therefore be taken into account in the numerical simulations (González et al. 2006, Busquet et al. 2007, González et al. 2009).

The cylindrical targets are used only for spectroscopic purposes. Two targets, G32 and G35, are in aluminium, with a length of 2.8 mm. The last target, G50, is in gold, and is also 2.8 mm long. They are closed on the backside by an "XUV window" for backside spectroscopic measurements, in the direction of the shock propagation.

All targets are closed on the backside by various "XUV windows": Al (200 nm), SiC (100 nm) and Si$_3$N$_4$ (500 nm). They are closed on the top by a composite film, made from polystyrene (10 μm) with gold coating (0.5 μm thick). The PALS laser, focalized on the foil (hereafter called the piston), ablates the polystyrene. The piston is accelerated by rocket effect and launches a strong, radiating shock in the Xenon gas. The gold foil aims at preventing the X-ray radiation emitted in the ablated CH plasma corona to heat the gas inside the tube.

The characteristics of the different targets are summarized in Table 1.

| Target | Section form | Transverse size (mm) | Length (mm) | Walls | Gas, Pressure in bar | Back side XUV window | Shot energy |
|---|---|---|---|---|---|---|---|
| F7 | Square | 0.7 x 0.7 | 6 | Gilt Al & Gilt BK7 | Xe, 0.1 | SiC 100 nm | 166 J |
| F61 | Square | 1.2 x 1.2 | 6 | Al & aluminized BK7 | Xe, 0.1 | $Si_3N_4$ 500 nm | 151 J |
| G32 | Circular | 0.7 | 2.8 | Al | Xe, 0.1 | Al 200 nm | 168 J |
| G35 | Circular | 0.7 | 2.8 | Al | Xe, 0.1 | SiC 100 nm | ~ 170 J |
| G50 | Circular | 0.7 | 2.8 | Gold | Xe+Air (50%+50%), 0.17 | Al 200 nm | 171 J |

Table 1: Summary of the characteristics of the different targets.

### 3 – LONGITUDINAL INTERFEROMETRY

Longitudinal interferometry traces the instantaneous variation of the optical index $n(x=0, z, t, \lambda)$, at the wavelength $\lambda$ of the probe laser, versus the position z on the cell axis (i.e. x=0), along the direction z of the shock propagation, and integrated over the path of the probe laser (y direction). The variations of the plasma optical index, $n(\lambda)-1$, are usually attributed to free electrons, and related to the electron density $N_e/N_{c,\lambda}$, where $N_{c,\lambda}$ is the critical density at the wavelength $\lambda$ of the probe laser,

$$n(\lambda) = [1-N_e/N_{c,\lambda}]^{1/2} \approx 1 - 0.5\, N_e/N_{c,\lambda}$$

(1)

The interferometric measurements thus provide the instantaneous electron density at the centre (x=0) of the tube, as function of the position z, and averaged over the transverse section d of the plasma

$$<N_e(z,t)>_{x=0} = \int_0^d N_e(x=0, y; z, t)\, dy/d$$

(2)

**3.1 General features**

We present in this section the results obtained with the longitudinal interferometric diagnostic for target F7, filled with Xe at 0.1 bar. The post-mortem target analysis indicates that the PALS laser impact was satisfactorily centered on the target. The X-ray pinhole record shows a smooth distribution of the laser intensity. The raw image is presented without *(a)* and with *(b)* annotations in Figure 3. The vertical axis is the time t (100 ns, 0.2 ns/pixel). Time zero at the bottom of the figure corresponds to the laser arrival on the target. The horizontal axis is the position z along the tube centre, which is imaged on the slit (6 mm, 12 μm /pixel in z). The time resolution is shorter than 1 ns. Due to the delayed synchronization, we did not record the fringes before the time of arrival of the PALS laser pulse on the gilt plastic foil, which is located on the left in Fig. 3. The time of laser arrival may be recognized in the small fringe deviations, which are visible at the bottom of the figure (t < 4 ns). These deviations may be attributed to electromagnetic perturbations in the streak camera or to spurious heating of the target associated to unshielded stray light from the PALS laser. However, these deviations are associated in the worst case with low electron densities, and may be discarded in the interpretation of the interferogram.

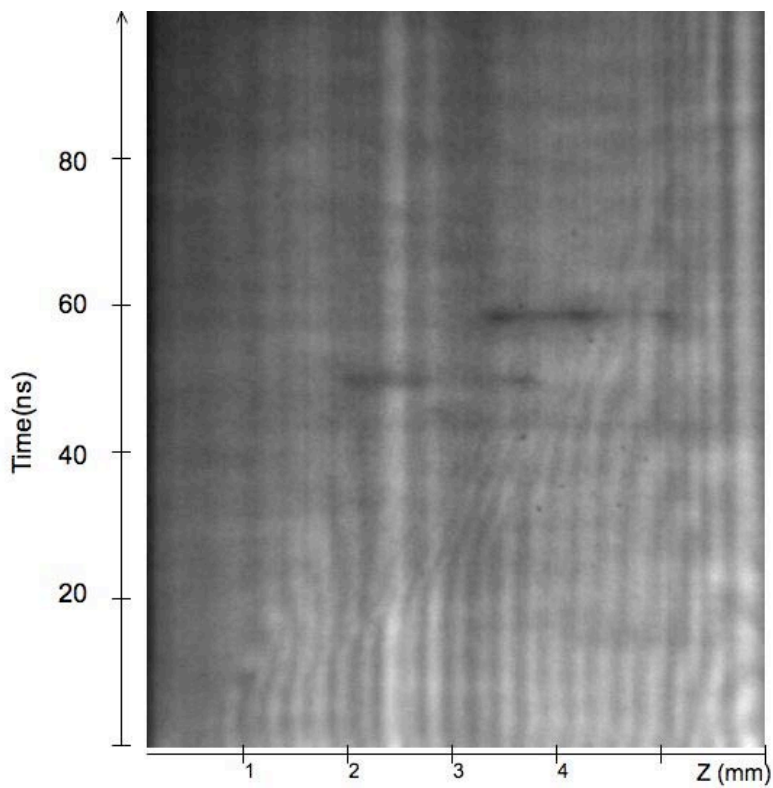 (a)

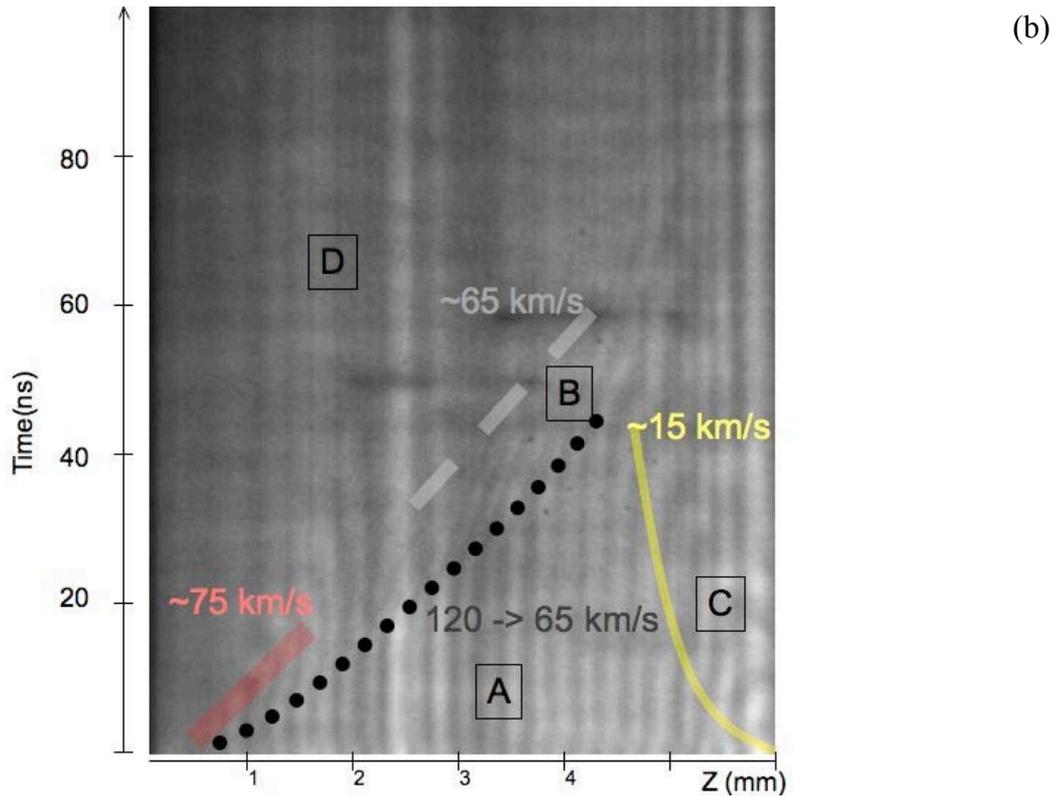

Figure 3: Images of the interferometric longitudinal diagnostic for the target F7 at 0.1 bar. Time along the vertical axis is in ns. The horizontal axis reports the position z in mm of the shock propagation from left to right. The laser pulse comes from the left. Superimposed lines in the right panel give an approximate guide to the different time and space domains discussed in the text.

The interferogram presents four distinct zones of moving fringes (see Figure 3b):
- the first zone – **A** – corresponds to t < 50 ns and z < 4 mm. In this zone, the fringes are unperturbed (i.e., vertical) until they start to bend towards the right and to become narrower. The location of this bending (corresponding to about ¼ of fringe deviation) is qualitatively reported in Fig. 3b with the dotted black line. The bending is the signature of a positive gradient of the electron density column, and is the characteristic signature of the ionisation produced by the radiative precursor of the shock (Fleury et al. 2002, Bouquet et al. 2004), which is visible in second zone -**B**. The velocity of the ionisation front decreases from 120 to 65 km/s as the shock propagates from left to right.

For z > 3 mm, and time greater than 25 ns, the narrow bent fringes of the radiative precursor disappear. The location of this event is approximately reported by a full thick grey line, and corresponds to a velocity of 65 ± 7 km/s. For z < 1.8 mm, t < 20 ns, the fringes, originally vertical, bend towards the right, and then towards the left. The inflexion points are located over the full thick red line, and are associated to a velocity of 75 ± 13 km/s. Close to this inflexion point, some fringes present a small discontinuity.

The third zone – **D** – in the upper part of the image is located above the two thick lines. Whereas the interface between the piston and the gas, as inferred from MULTI 1D (Ramis et al., 1988) simulations for typical PALS laser conditions (1.5 $10^{14}$ W/cm$^2$), is close to these two thick lines (see Fig. 4), the presence of fringes in the left part of **D** (i.e., around z ~ 1–1.5 mm, t ~ 5 – 15 ns) just above the thick lines, is not expected from these simulations and will require separate studies. Note that they are also seen at the canal centre up to 20 ns after the shock front in the transverse interferogram performed with larger cell sections (§4). For z larger than 1.5 mm and times longer than 15 ns, the fringes behaviour and their interpretation is similar to previous studies (Bouquet et al. 2004). Hereafter, we shall mainly focus the discussion on this part of the interferogram.

Finally, the last feature of the interferogram – **C** – , located towards the end of the tube (z > 4.5 mm), is identified by the full yellow line on the right-hand side of the figure. This feature corresponds to the development of a reverse ionization wave expanding from the heated backside of the tube (z = 6 mm) at a velocity of about 15–20 km/s. The interaction zone between this ionization wave and the radiative precursor in Xenon is located between 4 and 5 mm and t < 50 ns. The development of this plasma from the end of the tube explains the absence of record on the XUV spectrometer from the backside from the tube. The plasma expansion from the end of the square channel may be due to gas preheating which may originate either from particles (fast electrons, which seems however unlikely for the moderate laser irradiance, $10^{14}$ W/cm$^2$, see for instance, Drake et al. 1984, Batani et al. 2003) or from parasitic laser irradiation. A possible explanation of this expansion may come from the direct irradiation of the backside of the tube by the residual IR PALS laser beam (few J). This IR beam is less focused than the main beam at 438 nm and, therefore, it is not perfectly screened off by the piston foil. Consequently, this beam may reach the end of the square tube before being blocked by the socket. In the following, we shall thus restrict the interpretation of the interferometric diagnostic to times shorter than 50 ns and to distances smaller than 4 mm that are not impacted by the effect of the reverse plasma expanding from the backside. We stress however that the pre-heating does not influence the XUV diagnostics realized in the blind targets.

## 3.2 Shock dynamics and structure

Let us follow the fringes displacement with time. We shall assume that the dephasing is mainly due to free-free absorption, which relates the average transverse electron density $<N_e(z,t)>_{x=0}$ to the phase shift over the section d of the tube. The phase shift is given by:

$$\varphi(z,t) = \frac{2\pi}{\lambda} \int_0^d \left( 1 - \sqrt{1 - \frac{N_e(x=0,y,z,t)}{N_{c,\lambda}}} \right) dy \approx \frac{\pi d}{\lambda} \frac{<N_e(z,t)>_{x=0}}{N_{c,\lambda}},$$

(3)

where $\lambda = 527$ nm is the wavelength of the probe laser, $<N_e(z,t)>_{x=0}$ is given by Eq. (2), $N_{c,\lambda} = 4\ 10^{21}$ cm$^{-3}$ is the critical density at 527 nm, and d = 0.7 mm is the tube section.

We disregard the first fast dephasing discussed in the previous section. The fringe maxima and the reconstructed isocontours of $<N_e(z,t)>_{x=0}$ are reported in Figure 4 for densities between $2\ 10^{17}$ and $10^{19}$ cm$^{-3}$. The distance between two fringes corresponds to about 17 pixels and a density of $6\ 10^{18}$ cm$^{-3}$.

Each isocontour profile is concave, and tends approximately towards the same linear slope between 20 ns and 40 ns, corresponding to a velocity $v_s$ of 63 km/s (±10%), similar to the velocity of the gold-xenon interface calculated with MULTI 1D (not observed in the image, but reported for reference as the black dashed line). In this time interval, the residual extension of the radiative precursor is small, and the shock approaches the quasi-stationary limit.

Previous studies (Gonzalez et al. 2006, Busquet et al. 2007, Gonzalez et al. 2009) have already shown that the lateral radiative losses play an important role in the shock structure and dynamics. In particular, the precursor is curved and less extended in presence of radiative losses. The curvature is associated with the presence of colder layers close to the tube walls. Moreover, it was shown that the time to reach the stationary limit decreases with radiative losses.

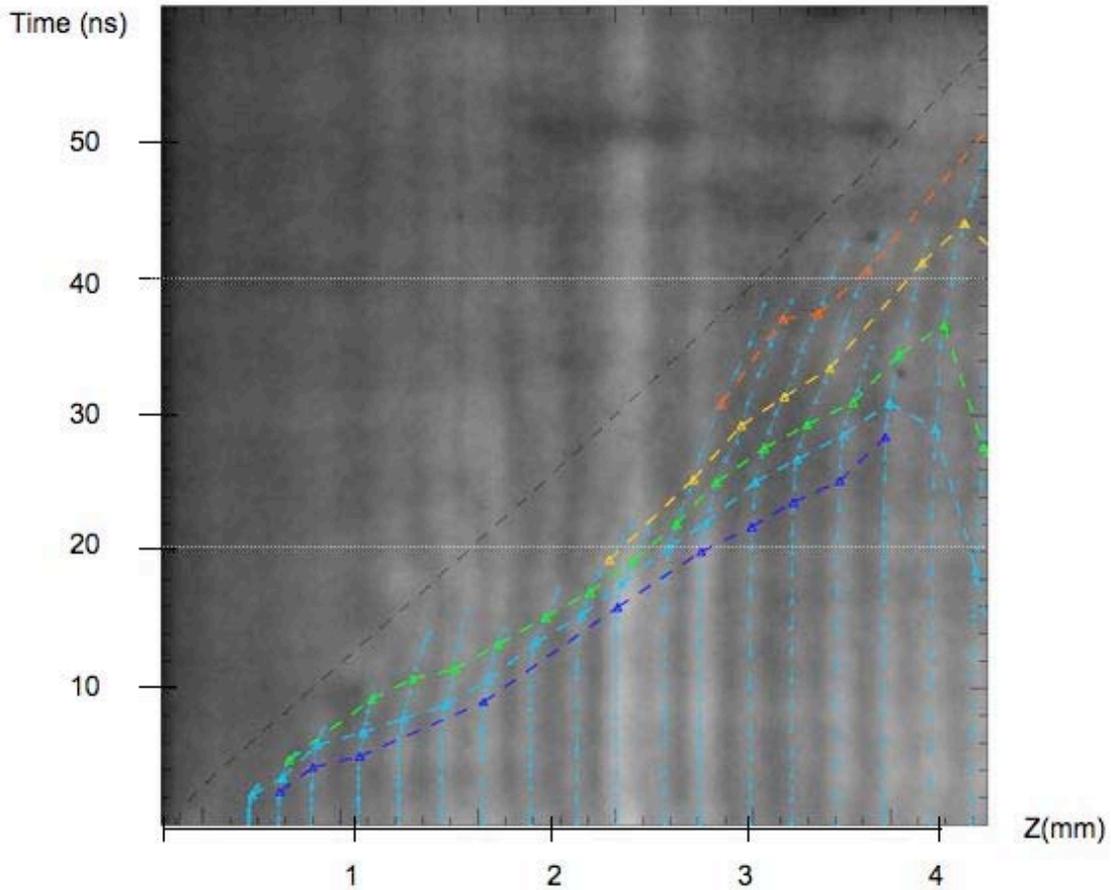

Figure 4: Section of the interferogram plotted on Fig. 3, showing the fringe maxima in dashed blue lines. Time is on vertical axis (60 ns), and position z in horizontal axis (4.2 mm). The symbols on the horizontal axis at t=0 report the location of the unperturbed fringes maxima from a previous record. The square symbols on the fringes locate isocontours of <$N_e$> for 2. $10^{17}$ (dark blue), 7 $10^{17}$ (blue), 2. $10^{18}$ (green), 5 $10^{18}$ (yellow) and $10^{19}$ cm$^{-3}$ (red). The position of the gold-xenon interface computed with MULTI 1D, not visible on the record, is reported for reference (black dashed line).

We have studied numerically the properties of a radiative shock in a 2D geometry for different values of the wall radiative losses and for a constant piston velocity of 63 km/s. The choice of this velocity follows from the measured velocity of the isodensity contours between 20 ns and 40 ns. The simulations were performed with the HERACLES 2D radiation hydrodynamic code (Gonzalez et al. 2007), which includes radiation transport, Super Transition Array opacity (Bar-Shalom et al. 1989) and an equation of state from the OPA-CS hydrogenic model (Michaut et al. 2004). Figure 5 shows the calculated and the measured average transverse electron density in the precursor at two different times (20 ns and 40 ns), for two values of the radiative losses (20% and 40%) at the gilt walls of the tube. The experimental results indicate that the radiative losses are of the order of 30 to 40 %. These losses measured in a tube with gold-coated walls are smaller than those measured in earlier studies (60%) with tubes of aluminized glass (Gonzalez et al. 2006, Busquet et al. 2007).

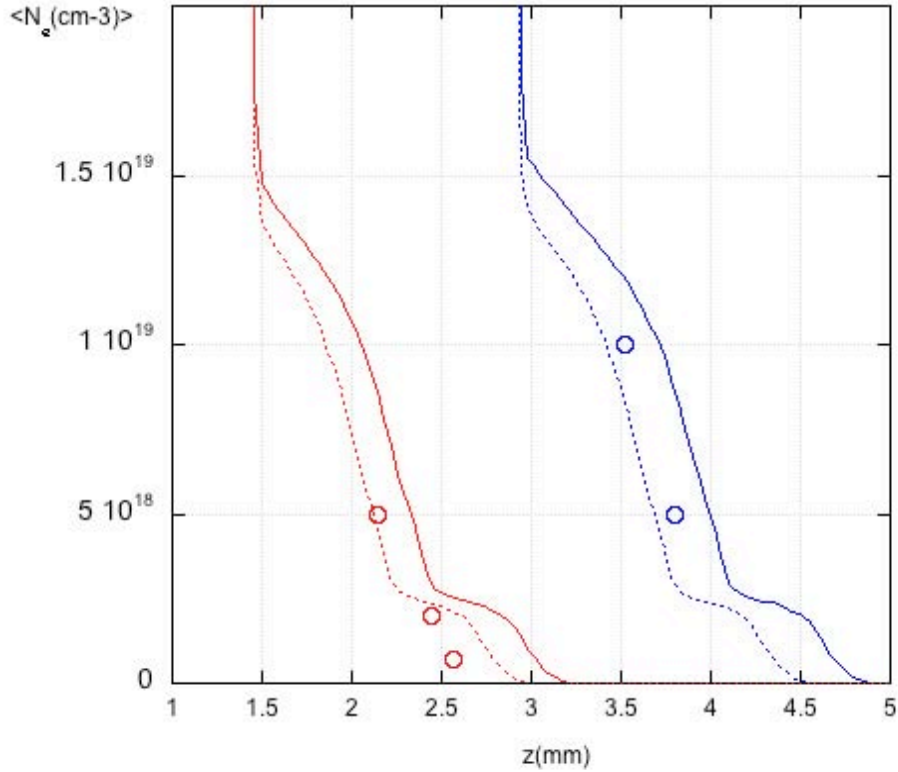

Figure 5: Transverse electron density $<N_e(z,t)>_{x=0}$ at two times (20 ns in red and 40 ns in blue): experimental value (circles) and calculated electron densities with HERACLES for two values of the radiative losses (40 % in dashed; 20 % in full line). Due to measurement uncertainties, from the interferogram, the point at $2\ 10^{17}$ cm$^{-3}$ at 20 ns has been discarded. The precision of the measurement is ~ 0.1 mm

## 4 – TRANSVERSE INTERFEROMETRY

Longitudinal interferometry probes the time variation of the averaged transverse electron density $<N_e(z,t)>_{x=0}$ along the canal centre (x=0), versus the direction z of the shock propagation. This diagnostic allows the study of the propagation of the ionisation front in the tube, but it does not give any direct information about the 2D structure of the radiative shock. To investigate the 2D structure, we performed a few transverse interferometric studies where the slit of the streak camera records the image, at a fixed distance $z_0$ from the piston, in the direction x perpendicular to the shock propagation. Transverse interferometry probes the time variation $<N_e(x,t)>_{z=z_0}$ of the electron density averaged over the transverse direction y, for a fixed value of $z_0$ along the tube.

For this study we have used a different target (F61, see Table 1). Its BK7 windows have an aluminium coating and the other faces of the tube are massive aluminium. The section of the cell, 1.2 x 1.2 mm$^2$ is larger than the nominal focal spot (~ 0.55 mm diameter) and the tube length is 6 mm. The target is filled with xenon at 0.1 bar, and we imaged the section of the tube at $z_0$ ~ 3 ± 0.25 mm.

The interferogram is displayed in Figure 6. The horizontal axis corresponds to the transverse direction x (1.2 mm) and the vertical axis is the time (100 ns). The position of the time arrival of the laser pulse is reported in the figure by a dashed line. The figure shows that the shock section varies from less than 0.7 mm at 36 ns (precursor) to a section smaller than 1 mm at 42 ns. Although we did not see any evidence of the laser beam in the post-mortem analysis of the target, the record shows that the shot was not perfectly centred in the channel. Following the fringes vertically from top to bottom in the figure, after the time of shock arrival, we point to a small bending (black arrow) to the right at ~ 33 ns (precursor), reaching up to one fringe shift at ~ 40 ns in the half right part of the image. After 40 ns the shock starts to be more opaque, especially on the left part of the figure. The location of the densest part of the shock is visible on the right part of the interferogram by this darkening and fringe discontinuity (large arrow). The corresponding time of arrival is ~ 42 ns, yielding a shock velocity of 71 ± 6 km/s. The precursor duration is ~ 7 ns, which is smaller than the duration that can be derived from Fig. 4 at z = 3 mm (which exceeds the 14 ns measured between 7 $10^{17}$ and $10^{19}$ cm$^{-3}$). The fringes remain visible up to the canal centre and bend towards the left after the shock, where their topology becomes irregular (bottom of the figure). The precursor (t ~ 33–40 ns) is smoothly bent as a consequence of both hydrodynamical expansion and radiation effects.

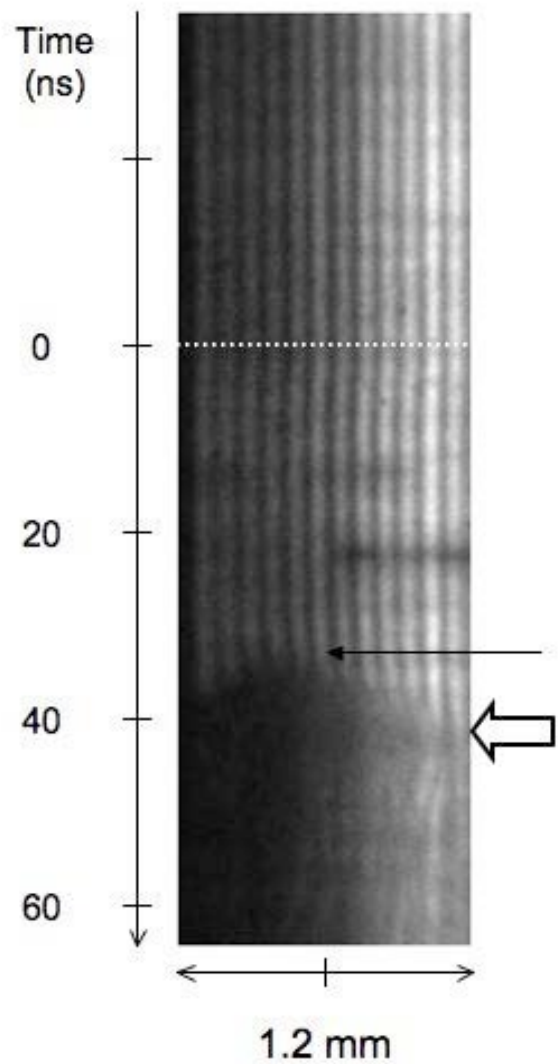

Figure 6: Transverse interferogram of the shock (target F61), versus time t (vertical axis) and transverse direction x (horizontal axis) for z0 ~ 3 mm along the cell axis. Scales are 100 ns for the time and 1.2 mm for x. The dashed line corresponds to the time of laser pulse arrival. Contrarily to other figures, time increases from top to bottom.

## 5 – REAR SIDE XUV SPECTROSCOPY

Whereas interferometry probes the plasma electron density, the plasma ionization and temperature can be investigated using spectroscopy. Photometric and spectroscopic signatures of the radiative shocks provide some critical tests to radiation hydrodynamic codes. In the present case of a radiative shock, the opacity variation along the shock structure is key to understand the line spectrum formation. As pointed out already by Gonzalez et al. (2006), the average opacity of the dense part of the shock is large. The opacity is small in the heated precursor, but large again in the cold unshocked gas. Hence, the extreme ultraviolet (XUV) photons emerging from the shock front propagate almost freely up to the precursor front where they are reabsorbed. This process contributes to ionize the gas in front of the precursor and thus to increase its length. Qualitative synthetic spectrum calculations (Stehlé et al. 2009), carried out for shocks in similar conditions than the present experiment, show that the emerging flux varies strongly with time because of the intrinsic time variations of the shock conditions. Additionally, flux variations may also be expected as a result of the decreasing thickness of the cold absorbing gas between the shock and the rear face of the tube as the shock propagates in the tube.

We have carried out an initial investigation in order to check the feasibility of characterizing the radiative flux emerging from the back face of the tube. Since the maximum emission from the shock is expected to be in the XUV range, we have realized an experimental setup with a flat field XUV spectrometer, with time integration over 16 µs, to study the spectral signatures of a radiative shock in Xenon around $\lambda = 15$ nm for shock velocities of ~ 65 km/s. The recorded spectrum shown in Figure 7 extends from 12 nm to 23 nm. As an aluminium filter of 1.6 µm is used to protect the grating against debris from the target in addition to the backside XUV window, the flux at wavelengths shorter than 17.1 nm (L edge cutoff of aluminium) is severely reduced by the filter extinction. The experimental setup is described in §2, with the collection angle $\Omega$ on the CCD camera, estimated to $1.4 \cdot 10^{-4}$ rad, chosen to probe the central section of the tube.

For this spectroscopic study, we used blind targets either in aluminium (G32, G35) or in gold (G50). The inner diameter of the shock tube is 0.7 mm and the length is 2.8 mm. These 2 mm are followed by an expansion chamber (diameter 2 mm and length 0.1 mm ) to weaken the shock wave. An XUV filter is placed at the backside of the tube, and is made of sub-micron foils of SiC or Aluminium. The gas is either Xenon at 0.1 bar or a 50-50 mixture of air and Xenon at 0.17 bar (Table 1). The G35 shot (black) was poorly centred, and the recorded conditions therefore differ from the plasma conditions probed in the G32 shot.

We show the recorded spectra for the three different targets in Figure 7. These spectra present four main spectral features at ~17.2, 18.3, 19.2 and 20.0 nm. The two last lines, which are present in all the targets, are attributed to Xenon. The two other features are attributed to oxygen, with possible blending from aluminium in the case of G35 and G32) coming from the walls.

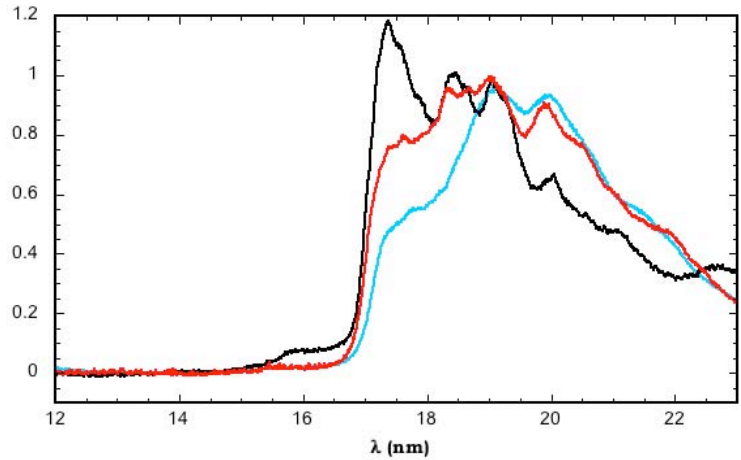

Figure 7: XUV spectra, normalized at 19 nm, for three different shots: black line: xenon at 0.1 bar in aluminium tube, and an aluminium XUV window (cell G35); red line: xenon at 0.1 bar in aluminium tube, and an SiC XUV window (cell G32); blue line: mixture of air and xenon (50%-50% in number of moles) at 0.17 bar in an gold tube (cell G50) and aluminium XUV window.

While the measured count numbers for the G32 and G35 targets filled with xenon (about 1000 counts/pixel) are in qualitative agreement with numerical predictions, the interpretation of the time-integrated spectra remains difficult and is beyond the scope of this paper. However, this preliminary experiment shows the feasibility of backside spectroscopic investigations. Future studies will benefit of improved time and spectral resolution and will require accurate monochromatic opacities for comprehensive spectrum analyses.

## 6 – CONCLUSIONS

In this article, we studied the propagation of a radiative shock in Xenon at 0.1 bar for times up to 50 ns and shock velocity of ~ 60 km/s. Using longitudinal interferometry diagnostics, we found small variations only of the velocities of unperturbed isodensity contours between 20 and 40 ns, indicating a limited residual development of the radiative precursor at these times, as expected from earlier studies (González et al, 2009). This result suggests that a quasi-stationary limit is approached at a shock velocity of 63 km/s (±10%). Our results are consistent with radiative losses at the gilt walls of the order of 30%. The reported values of the electron density are consistent with 2D radiative hydrodynamics simulations. Transverse interferometric measurements in larger channels (1.2x1.2 mm$^2$) show a small bending of the radiative precursor that is less extended in time than in smaller tubes. Detailed comparison between the experimental results and numerical simulations will be the topic of further study and will be published separately.

Despite the present lack of time resolution, our preliminary results on backside XUV spectroscopy are encouraging to prompt further spectroscopic studies of radiative shocks.

# 7 – ACKNOWLEDGMENTS

The authors acknowledge M. Busquet (LERMA) and F. Thais (CEA/IRAMIS) for their participation to the experiment and for informative discussions during the write up of the paper, O. Madouri (LPN), NCLA laboratory and V. Petitbon (IPN) for providing us with the SiC windows, the pinhole for the X-ray pinhole diagnostic, and the films of polystyrene. They also thank G. Loisel (CEA/IRFU) for his help during the experiment. We acknowledge financial supports from the Access to Research Infrastructures activity in the Sixth Framework Programme of the EU (contract RII3-CT-2003-506350 Laserlab Europe), from RTN JETSET (contract MRTN-CT-2004 005592), French National Programme of Stellar Physics (PNPS), and CNRS-PICS 4343. T.L. gratefully acknowledges financial support from Observatoire de Paris during the past several years. M.G. acknowledges the financial support provided by the Spanish Ministry of Science and Innovation through the Juan de la Cierva grant. This research was partially supported by the Czech Ministry of Education, Youth and Sports (project LC528). The work is part of the ANR-08-BLAN-0263-01 project.

# 8- REFERENCES

BAR-SHALOM, A., SHVARTS, D., OREG, J., GOLDSTEIN, W. H., ZIGLER, A. (1989) Super-transition-arrays-A model for the spectral analysis of hot dense plasma, *PHYS. REV. A* **40**, 3183- 3193.

BATANI, D*.,* STRATI, F., TELARO, B., LÖWER, T, HALL, T., BENUZZI-MOUNAIX, A., KOENIG, M., (2003), Production of high quality shocks for equation of state experiments*, Eur. Phys. J. D* **23**, 99–107.

BATANI, D, DEZULIAN, R, REDAELLI, R, BENOCCI, R, STABILE, H, CANOVA, F, DESAI, T, LUCCHINI, G, KROUSKY, E, MASEK, K, PFEIFER, M, SKALA, J, DUDZAK, R, RUS, B, ULLSCHMIED, J, MALKA, V, FAURE, J, KOENIG, M, LIMPOUCH, J, NAZAROV, W, PEPLER, D, NAGAI, K, NORIMATSU, T & NISHIMURA, H (2007), 'Recent experiments on the hydrodynamics of laser-produced plasmas conducted at the PALS laboratory', *Laser and Particle Beams*, vol. 25, no. 1, pp. 127-41.

BERTOUT, C. (1989) T Tauri stars - Wild as dust, *Annual Review of Astronomy and Astrophysics*, **27**, 351-395.

BOUQUET, S., STEHLÉ, C., KOENIG, M., CHIÈZE, J.P., BENUZZI-MOUNAIX, A., BATANI, D., LEYGNAC, S., FLEURY, X., MERDJI, H., MICHAUT, C., THAIS, F., GRANDJOUAN, N., HALL, T., HENRY, E., MALKA, V., LAFON, J.P.J (2004), Observations of laser driven supercritical radiative shock precursors, *Phys. Rev. Lett.* **92** 5001.

BOUVIER, J., ALENCAR, S.H.P., HARRIES, T.J., JOHNS-KRULL, C.M., ROMANOVA, M.M. (2007), Magnetospheric Accretion in Classical T Tauri Stars, in Protostars and Planets V, B. Reipurth, D. Jewitt, and K. Keil (eds.), University of Arizona Press, Tucson, p. 479-494, (arXiv:astro-ph/0603498).

BOZIER, J. C., THIELL, G., LEBRETON, J. P., AZRA, S., DECROISETTE, M., SCHIRMANN, D., (1986). Experimental-observation of a radiative wave generated in xenon by a laser-driven supercritical shock, *Phys. Rev. Lett.,* **57**, 1304.

BUSQUET M., AUDIT E. , GONZALEZ M., STEHLÉ C., THAIS F. , ACEF O., BAUDUIN D., BARROSO P., RUS B., KOZLOVA M., POLAN J., MOCEK T. , Effect of lateral radiative losses on radiative shock propagation, *High Energy Density Physics*, 3, 8-11 (2007).

DONATI, J. F., JARDINE, M.M., GREGORY, S.G., PETIT, P., PALETOU, F., BOUVIER,

J. , DOUGADOS, C., MENARD, F., CAMERON, A.C., HARRIES, T.J., HUSSAIN, G.A.J., UNRUH, Y. , MORIN, J., MARSDEN, S.C., MANSET, N., AURIERE, M.N CATALA, C., ALECIAN, E. (2008) Magnetospheric accretion on the T Tauri star BP Tauri, *Mon. Not. R. Astron. Soc.,* **386**, 1234-1251.

DRAKE, R.P., TURNER, R.E., LASINSKI, B.F., ESTABROK, K.G., CAMPBELL, E.M., WANG, C.L., PHILLION, D.W., WILLIAMS, E.A., KRUER, W.L., (1984), Efficient Raman Sidescatter and Hot –Electron Production in Laser-Plasma Interaction Experiments, Phys. Rev. Let. 53, 1739-1742.

FLEURY, X., BOUQUET, S., STEHLÉ, C., KOENIG, M., BATANI, D., BENUZZI-MOUNAIX, A., CHIÈZE, J.P., GRANDJOUAN, N., GRENIER, J., HALL, T. ,HENRY, E., LAFON, J.P.J, LEYGNAC, S., MALKA, V., MARCHET, B., MERDJI, H., MICHAUT, C., THAIS, F., (2002), A laser experiment for studying radiative shocks in astrophysics, *Laser Part. Beams,* **20,** 263.

GONZÁLEZ, M., AUDIT, E., HUYNH, P. (2007). HERACLES: a three dimensional radiation hydrodynamics code, *A&A,* **464**, 429-435.

GONZÁLEZ, M., STEHLÉ, C. , AUDIT, E. , BUSQUET, M., RUS, B., THAIS, F., ACEF, O. , BARROSO, P., BAR-SHALOM, A., BAUDUIN, D., KOZLOVA, M., LERY, T., MADOURI, A., MOCEK, T., POLAN, J. (2006) J. Astrophysical radiative shocks : from modelling to laboratory experiments, *Laser Part. Beams,* **24**, 535-545

GONZÁLEZ, M., AUDIT, E., STEHLÉ, C. (2009) 2D numerical study of the radiation influence on shock structure relevant to laboratory astrophysics, ,*A&A* **, 497**, 27-34

JUNGWIRTH, K. (2005). Recent highlights of the PALS research program. *Laser Part. Beams* **23**, 177–182.

KASPERCZUK, A, PISARCZYK, T, NICOLAI, PH, STENZ, CH, TIKHONCHUK, V, KALAL, M, ULLSCHMIED, J, KROUSKY, E, MASEK, K, PFEIFER, M, ROHLENA, K, SKALA, J, KLIR, D, KRAVARIK, J, KUBES, P & PISARCZYK, P (2009) 'Investigations of plasma jet interaction with ambient gases by multi-frame interferometric and X-ray pinhole camera systems', *Laser and Particle Beams*, vol. 27, no. 1, pp. 115-22

KISELEV YU, N. , NEMCHINOV I.V., SHUVALOV V.V (1991) , Mathematical modelling of the propagation of intensely radiating shock waves, Comput. Maths. Math. Phys., **31**, No.6, pp.87-101

KLOSNER, M.A, SILFVAST, W.T., (2000), Xenon-emission –spectra identification in 5-20 nm spectral region in highly ionized xenon capillary-discharge plasmas, *J. Opt. Soc.*

*Am. B*, 17, 1279-1290

MICHAUT, C., STEHLÉ, C., LEYGNAC, S., LANZ, T., BOIREAU, L., (2004), Jump conditions in hypersonic shocks. Quantitative effects of ionic excitation and radiation, *Eur. Phys. J. D*, **28,** 381-392

MIHALAS, D., MIHALAS, B. D. (1984). *Foundation of Radiation Hydrodynamics.*, Oxford University Press.

NILSEN, J. , JOHNSON, W. R. (2005), Plasma interferometry and how the bound-electron contribution can bend fringes in unexpected ways, *Applied Optics*, **44**, 7295

RAGA, A.C., MELLEMA, G., ARTHUR, S.J., BINETTE, L., FERRUIT, P., STEFFEN, W. (1999), 3D Transfer of the Diffuse Ionizing Radiation in ISM Flows and the Preionization of a Herbig-Haro Working Surface, *Rev. Mex. Astron. Astrof.*, **35,** 123

RAMIS, R., SCHMALZ, R. & MEYER-TER-VEHN, J. (1988). MULTI – a computer code for onedimensional multigroup radiation hydrodynamics, *Comp. Phys. Comm.*, **49**, 475.

REIGHARD, A.B., DRAKE, R.P., DANNENBERG, K., PERRY, T.S., ROBEY, H.A., REMINGTON, B.A., WALLACE, R.J., RYUTOV, D.D., GREENOUGH, J., KNAUER, J., BOELHY, T., BOUQUET, S., CALDER, A., ROSNER, R., FRYXELL, B., ARNETT, D., KOENIG, M. (2005), Collapsing radiative shocks in argon gas on the omega laser, ApSS, **298**, Issue 1-2.

REIGHARD, A.B., DRAKE, R.P., MUCINO, J.E., KNAUER, J.P., BUSQUET, M. (2007), Planar Radiative shock experiments and their comparisons with simulations, Phys. Plasmas **14**, 056504.

RODRIGUEZ, R, FLORIDO, R, GLL, JM, RUBIANO, JG, MARTEL, P & MINGUEZ, E (2008), 'RAPCAL code: A flexible package to compute radiative properties for optically thin and thick low and high-Z plasmas in a wide range of density and temperature', *Laser and Particle Beams*, vol. 26, no. 3, pp. 433-48.

STEHLE, C, GONZALEZ, M., AUDIT, E., Radiative shocks in the context of Young Stellar Objects : a combined analysis from experiments and simulations, proceedings of the conference « Protostellar jets in context », Rhodos, APSS Proceedings, Springer Verlag, K. Tsinganos, M. Stute, T. Ray, Ed. in press (2009), p117-121.

ZELDOVICH, Y.B. , RAISER, Y.P. , (1966) *Physics of shock waves and high temperature hydrodynamic phenomena*, New York: Academic